# Cyclic spectra for wavelength-routed optical networks

Bill Corcoran[1,2,*], Zihan Geng[1], Valery Rozental[1], and Arthur J. Lowery[1,2]

[1]*Monash Electro-Photonics Laboratory, Department of Electrical and Computer Systems Engineering, Monash University, VIC 3800, Australia*
[2]*Centre for Ultrahigh-bandwidth Devices for Optical Systems, Australia*
[*]*Corresponding author: bill.corcoran@monash.edu*

**We propose occupying guard-bands in closely-spaced WDM systems with redundant signal spectral components, to increase tolerance to frequency misalignment and channel shaping from multiplexing elements. By cyclically repeating the spectrum of a modulated signal, we show improved tolerance to impairments due to add/drop multiplexing with a commercial wavelength selective switch, in systems using 5-20% guard bands on a 50-GHz DWDM grid.** © 2017 Optical Society of America

*OCIS codes:* some codes here.

http://dx.doi.org/xxxx

Optical routing of signals using reconfigurable wavelength selective switches in an optical fiber communications network provides redundancy in case of natural disasters or cable breaks [1], lower cost switching in photonic networks [OFTVIB Ch15], and low energy consumption [2]. Wavelength division multiplexing (WDM) is used to channelize the >4 THz bandwidth available in optical fiber systems employing erbium doped fiber amplifiers (EDFAs), commonly with channels equally separated by 50 or 100 GHz. The limited resolution of optical filters provided by WDM multiplexers, such as reconfigurable wavelength selective switches [3], leads to the use of reserved spectrum for guard-bands between channels, leaving somewhere between 36-44% of the available optical spectrum underutilized [4]. Spectral shaping of optical signals provides a method to efficiently use channel bandwidth. Spectral efficiency can be improved through duo-binary pulse shaping [5], or more complex schemes [6], albeit at the cost of reduced sensitivity. The pulse shaping used to create optical super-channels that have close-to-rectangular signal spectra can restrict the signal bandwidth to close to the signal baud rate [7, 8], ideally without a sensitivity penalty. While there are several proposed solutions to allow for signal bands that are spaced very closely (< 5% guard band) in the spectral domain to be independently routed [9–12], commercially available wavelength selective switch (WSS) devices do not have sufficient wavelength resolution for this task. Intuitively, rectangular optical spectra allow more efficient use of channel bandwidth as defined by optical filtering for WDM multiplexing (mux) and/or demultiplexing (demux) than the sinc-shaped spectra of non-return-to-zero signals that are commonly used in WDM networks.

Spectrally shaped signals are therefore interesting in systems using moderate guard bands (e.g. 5-33%), to allow for efficient use of installed fiber optic communications infrastructure.

In Nyquist WDM [7], rectangular spectra are defined for each wavelength through filtering, generally using digital filters before modulating onto an optical carrier. While ideally a Nyquist-shaped signal is inter-symbol-interference (ISI) free, spectral shaping of such a signal, by an optical filter for example, can produce impairments resulting in reduced receiver sensitivity [13]. This leaves Nyquist WDM bands vulnerable to impairments from optical add/drop multiplexing in a wavelength-routed system.

Here, we introduce the concept of employing cyclic spectra to provide robustness to optical filtering. The addition of repeated spectral components to an Nyquist-shaped signal provide redundancy that can then be exploited by adaptive equalizers at the receiver side to reduce the impact of misalignment to the multiplexing filters that define the optical channel, and to pulse shaping by repeated passes through WDM mux/demux stages. We show these advantages through experiment, in systems employing guard bands of 5-20%, running 40-47.5 Gbd signals on a 50 GHz grid with add drop-multiplexing emulated using a commercially available liquid-crystal-on-silicon WSS. This demonstration shows that by filling guard-bands with redundant data, transmission performance can be improved in wavelength routed optical networks.

Nyquist WDM signals are commonly produced by passing a signal through a root-raised cosine filter at the transmitter at the beginning of an optical link, then matched filtered at the receiver end of the link. This filter is commonly set to have a bandwidth matching the baud (symbol) rate of the signal, restricting the signal to its minimum ISI-free 'Nyquist' bandwidth. In order to generate a spectrally flat signal, the filtered signal should be locally flat (or 'white') within the filter bandwidth (as in Fig. 1). This can be achieved by simply interleaving zeros into the signal in the digital (sampled) time domain when the sample rate in the digital domain is greater than 1 Sa/Symb, i.e. by generating a return-to-zero signal. If the sampling rate is an integer multiple of the baud rate, zero padding in this manner generates a white spectrum across the whole digitally defined band. If such a white spectrum is shaped with a rectangular filter with the same bandwidth as the signal baud rate, the signal spectrum contains unique components along its extent. If a filter with a bandwidth greater than the baud rate is used, there are portions of the spectrum that are repeated at the edges, so that the spectrum is cyclic from one edge to the other. Hence, in this paper, where we define flat spectra with excess bandwidth we denote these signals as having cyclic spectra. This principle is shown in Fig.1.

The concept of using excess bandwidth to improve fiber communication system performance has been investigated in several contexts. Using excess bandwidths of over 100%, either through use of NRZ signaling [14], or through complex pulse shaping [15], has been shown to improve tolerance to optical nonlinear distortions in transmission. The advantage of having excess signal bandwidth in systems with WSS nodes has been briefly explored in a previous simulation-based investigation [13], where a comparison between 10% and 100% roll-off RRC shaping was performed, with the excess bandwidth associated with the 100% roll-off seemingly improving tolerance to filtering impairments. This observation was recently supported by an experimental investigation where bands of a 'folded OFDM' signal were more tolerant to add/drop multiplexing when the signal spectrum had redundant components [Corcoran]. Moreover, it has been shown that filtering a cyclic signal spectrum anywhere over its extent can allow for recovery of the modulated signal [16], which suggests that signals with cyclic spectra could be tolerant to mux/demux filter-signal frequency misalignment. Here, signals at line rates 40-47.5 Gbd, shaped with 1% roll-off RRC filters are compared, with the filter bandwidth set either to the signal baud rate to provide no excess bandwidth (i.e. forming Nyquist-shaped signals), or to 50 GHz to generate signals with cyclic spectral components (i.e. with excess bandwidths of 5-20%). We choose these values for baud rate not to match any required single channel bit rate, but to explore a region of spectral utilization beyoned that used in current systems (i.e. with guard bands occupying < 36% of the spectrum).

The transmitter and receiver DSP used to generate and receive both Nyquist-shaped and cyclic spectra signals is as follows. The modulated data is interleaved with zeros to provide a 50% duty cycle RZ signal, at 2 Sa/Symb. QPSK is used as a modulation format. This signal is then filtered by a 1% roll-off RRC filter. In the case of cyclic spectra, the signal is re-sampled to $2 \times 50/BR$ Sa/Symb before RRC filtering (where $BR$ is the signal baud rate in Gbd), generating a signal with excess bandwidth of $(50/BR) \times 100\%$. This signal is then re-sampled to match the transmitter digital-to-analog converter (DAC) rate of 90 GSa/s. At the receiver side, the received waveforms are re-sampled to 2 Sa/Symb, before residual-carrier frequency offset estimation, overlap-add dispersion compensation, pattern synchronization, and dynamic equalization. The dynamic equalizer operates in two stages. First, a training-based least-mean-squares (LMS) equalizer provides tap weights to initialize a blind constant modulus algorithm (CMA) equalizer, both with 81 taps. Phase estimation is performed using a training-aided maximum likelihood algorithm before bit-error rate calculation. The experimental set-up is shown in Fig. 2. Four external cavity lasers (ECLs) are set spaced by 50 GHz around a center frequency of 193.075 THz (1552.7 nm). Two pairs of waves separated by 100 GHz are sent to separate complex Mach-Zehnder modulators (CMZMs) driven by a 90-GSa/s (32-GHz electrical bandwidth) arbitrary waveform generator (AWG), then recombined to generate a four-channel band with an odd-even channel structure. Polarization multiplexing emulation through delay de-correlation is performed before optical noise loading. The signal can then be passed either directly to the receiver, or to a recirculating loop set-up consisting of a 80-km span of standard single mode fiber (SSMF) and an add/drop multiplexing emulation node. The node consists of a liquid-crystal-on-silicon WSS, with a 50-GHz bandwidth drop filter defined on a 50-GHz grid. The output of the 'drop' port is delay decorrelated from the output of the 'express' port, and the two outputs are combined with a 50/50 coupler. This emulates a worse case scenario for optical routing, where each node is set to filter the target channel. The receiver consists of a pre-filter before an integrated coherent receiver with outputs collected and digitized by a 80-GSa/s realtime oscilloscope. Receiver DSP is run off-line.

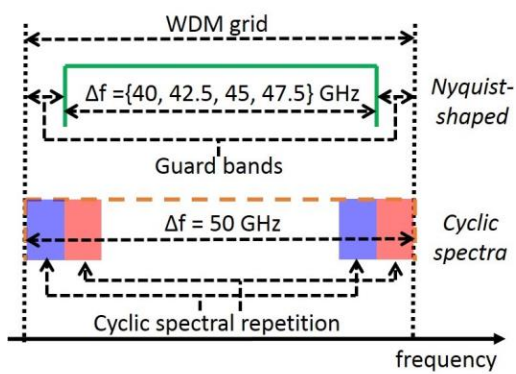

**Fig. 1.** Description of signal generation through spectral shaping of a zero-interleaved data. With 50 GHz channels, Nyquistshaped signal spectra occupy a bandwidth equal to their baud rate, while cyclic spectra signals occupy the full 50 GHz bandwidth with cyclically repeated spectral components.

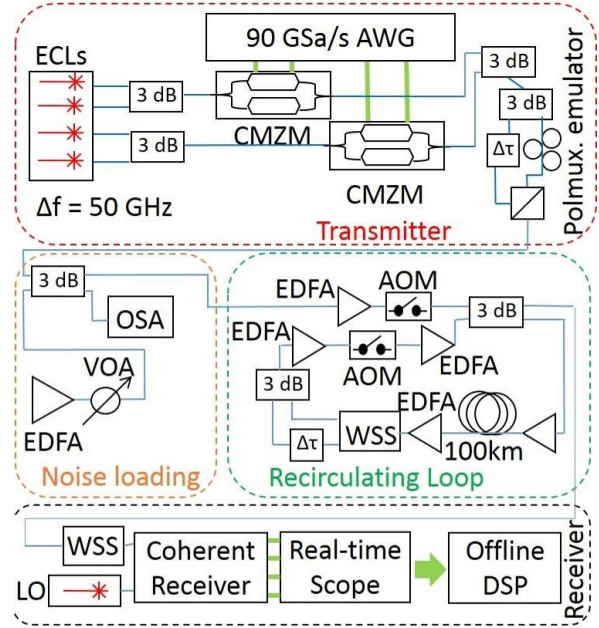

**Fig. 2.** Experimental set-up. Δτ: optical delay; OSA: Optical spectrum analyzer; AOM: Acousto-optic modulator; VOA: Variable optical attenuator; LO: Local oscillator

The Nyquist-shaped and cyclic-spectra signals are first compared in a back-to-back configuration with optical noise loading, to ensure that both types of signals are generated with comparable quality. Fig. 3 plots the back-to-back performance of both the Nyquist-shaped and cyclic spectra signals, and shows that both types of signals perform similarly against OSNR, indicating that both Nyquist-shaped and cyclic spectra signals are generated with similar quality. This is illustrated by the match in required OSNR for each baud rate, defined at the 7% ITU G.975.1 hard-decision forward error correction (HD-FEC) threshold ($Q^2$ = 8.56 dB). The degradation in required OSNR scales as expected as the baud rate is increased
If these results are interpreted slightly differently, so that performance is plotted against signal power spectral density over noise power spectral density (i.e against $S_{sig}/S_{ASE}$ in Fig. 3b), the cyclic spectra has better quality for the same spectral density ratio than the Nyquist-shaped signals, again illustrated by the inset of Fig. 3b which plots required spectral density ratio against baud rate.
In the case of the cyclic spectra, the signal is spread out over more bandwidth, intuitively resulting in an OSNR penalty counter to the measured result. This is a result of diversity gain enabled by the dynamic equalizer. The extra redundant spectral components in the cyclic spectrum are used by the dynamic equalizer, effectively resulting in a coherent addition of the extra spectral components with their central band counterparts to boost performance of the cyclic spectra signal (similar in concept to the spectral diversity scheme used in [17]). Supporting this, Fig. 3b shows 'enhanced' performance for the cyclic spectra signal from diversity gain, with the 'performance enhancement' (see inset in Fig. 3b) scaling as 50/$BR$ (i.e. 1, 0.7, 0.5 and 0.2 dB).

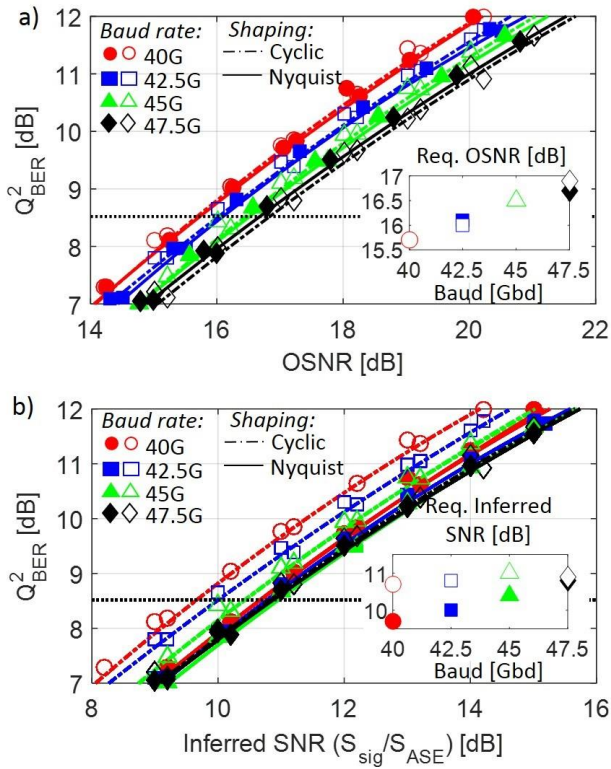

**Fig. 3.** Back-to-back performance of Nyquist-shaped (solid lines, close symbols) and cyclic spectra (dashed lines, open symbols) signals against a) OSNR and b) signal-to-noise power spectral density. Insets: Required OSNR or spectral-density-ratio for each signal tested.

In an optically routed link, there can be a significant mismatch (or detuning) between the center of the passband of add/drop multiplexers in the system and frequency of the optical signal. This mismatch can drift significantly over time, especially in systems where there is little or no active feedback to ensure device alignment to the set WDM grid. When passing through an optical filter or add/drop multiplexing stage, the main penalty on the filtered channel due to this mismatch will be due to channel shaping. As such, we detune the laser frequency of one of the central channels in the four-channel band off its nominal grid value of 193.1 THz, and pass the band though a single add/drop node.
Fig. 4 shows that detuning effects Nyquist-shaped signals much more than signals with cyclic spectra. This is most striking in the 40-Gbd case, where the initial $Q^2$ penalty due to filtering at 0-GHz detuning is negligible, but grows to 3.7 dB in at +5-GHz detuning. A similar trend is seen for the 42.5-Gbd and 45-Gbd signals, albeit with significant initial filtering penalties due to channel shaping after a single pass.
At 1-GHz detuning, a $Q^2$ performance advantage of 0.2, 1.2, 2.4 and 0.7 dB for the 40-, 42.5-, 45- and 47.5-Gbd signals, respectively. At 5-GHz detuning, this advantage changes to 3.7, 3.4, 2.5 and 0.6 dB for the 40-, 42.5-, 45- and 47.5-Gbd signals, respectively. As the baud rate is increased, the advantage of using cyclic spectra decreases for large detunings, as there is less redundant data at the edges of the signal spectrum. This indicates that by filling a guard-band between channels with redundant data, it is possible to generate signals that are more robust to filter misalignment.
Even when the signal is well aligned to the center of an add/drop filter passband, over multiple add/drop nodes, channel shaping from multiple filters can degrade performance. To ascertain the performance of Nyquist-shaped and cyclic spec-tra signals through multiple add/drop nodes in a WDM link, both signals were passed multiple times through the emulated add/drop node in the recirculating loop. Either all channels are passed express (Fig. 5a) or a single channel centered at 193.1 THz is dropped and re-added while the other channels are passed express (Fig. 5b), in the same way as for Fig. 4. Here, the add/drop channel is tuned to the nominal center of the filter band. For the 'all-pass' configuration, both Nyquist-shaped and cyclic spectra signals have similar performance for each of the signal baud rates tested. This is consistent with the back-to-back results presented in Fig. 3a. With the add drop filter in place, we start to see penalties from channel shaping by the add/drop filter. In this case, the cyclic spectra outperform the Nyquist-shaped signals, by 1-2 dB in quality factor ($Q^2$), or allowing an extra node to be reached (i.e. performance above the 7% HD-FEC threshold) for each of the signal baud rates tested. This indicates that not only are the cyclic spectra signals tolerant to channel misalignment, but also more tolerant than Nyquist-shaped signals to channel shaping over multiple add/drop filter passes (e.g. through multiple ROADM nodes).

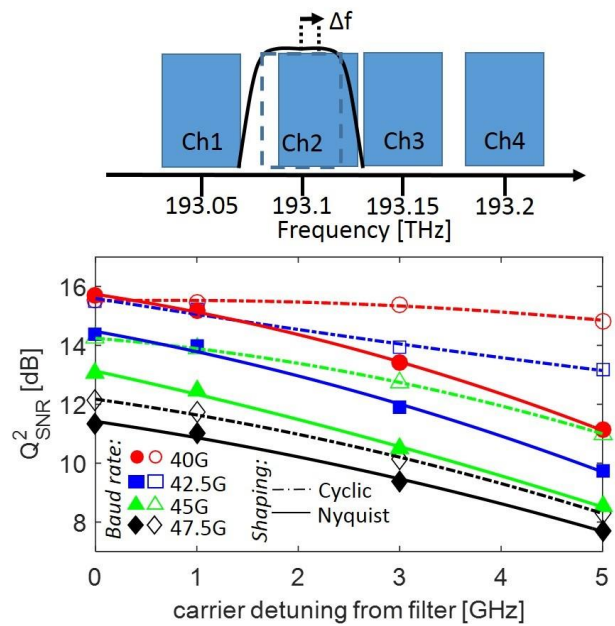

**Fig. 4.** Performance in the presence of signal frequency detuning off the set 50 GHz grid, after a single 80km span and add/drop node. The add/drop node set to add/drop mux a single channel at 193.1 THz, as illustrated by the top inset. Nyquist-shaped: Solid lines, closed symbols; cyclic spectra: dashed lines, closed symbols. Inset: illustration of detuning experiment.

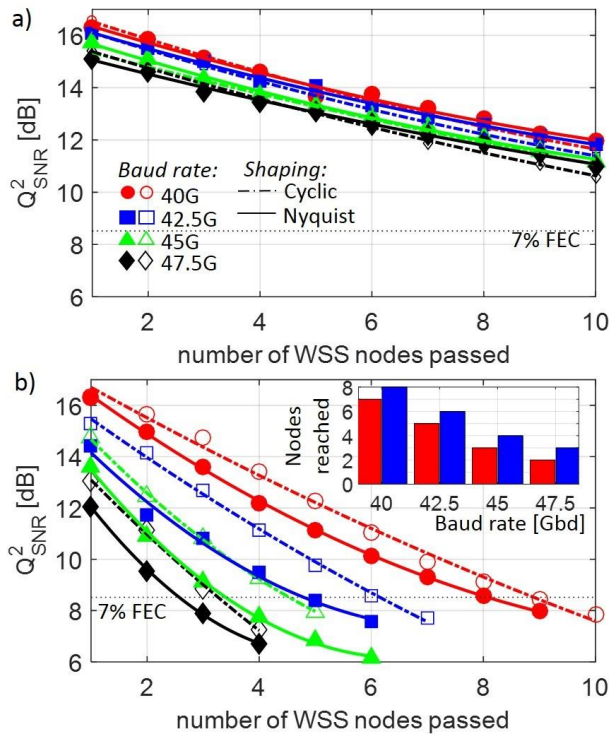

**Fig. 5.** Multi-pass performance of Nyquist-shaped and cyclic spectra signals through an emulated add/drop multiplexing node with the node set to a) 'all-pass' and b) to add/drop a single channel centered at 193.1 THz. Inset: Comparison of the number of nodes reached by each signal tested.

In summary, we have demonstrated that signals with cyclic spectra, carrying redundant data in bandwidth normally reserved as guard band, can improve tolerance to filter misalignment and channel shaping that can occur when using wavelength switching in WDM networks. The technique requires no additional receiver side digital signal processing steps, relying only on adaptive equalization to provide matched filtering. Moreover, the generation of cyclic spectra signals requires minor modification to digital signal processing at the transmitter side. This technique suggests that it is advantageous to use all of the available channel bandwidth in WDM networks, rather than the current approach of reserving unused spectral guard bands between wavelength channels.

Funding: Australian Research Council (ARC) (CE110001018, FL130100041)